\documentclass[9pt,conference]{IEEEtran} 
\newcommand{\IT}{\sffamily{STEALTH}}
\usepackage{mathpazo} 
\usepackage{framed} 
\usepackage{graphicx}
\usepackage{changepage}
\usepackage{cite}
\usepackage{color}

\definecolor{formalshade}{rgb}{0.93,0.93,0.93}

\definecolor{darkblue}{rgb}{0.2, 0.2, 0.2}

\newenvironment{formal}{%
  \MakeFramed{\advance\hsize-\width\FrameRestore}%
  \noindent\hspace{-1pt}% disable indenting first paragraph
  \begin{adjustwidth}{}{7pt}%
  \vspace{2pt}\vspace{2pt}%
}
{%
  \vspace{3pt}\end{adjustwidth}\endMakeFramed%
}

\begin{document}

\newcommand{\myfontsize}{\fontsize{16pt}{16pt}\selectfont}

\title{\textbf{\myfontsize Don't Lie to Me:    
  Avoiding Malicious Explanations with STEALTH}}

\author{ ~Lauren Alvarez, Tim Menzies \\
 North Carolina State University, USA\\
 lalvare@ncsu.edu, timm@ieee.org, \\
Submitted to IEEE Software, special issue on  XAI4SE,  December 2022}

\maketitle
\begin{abstract}
 
{\IT} is a method for using some  AI-generated  model,  without suffering
from malicious attacks (i.e. lying) or associated unfairness issues.
After recursively bi-clustering the data,  
{\IT}  system asks the AI model
a limited number of queries about class labels.
{\IT}  asks so few queries (1 per data cluster) that  
  malicious algorithms (a)~cannot
detect its operation, nor (b)~know when to lie.

In order to support open science practices, all our scripts and data are on-line at https://github.com/laurensalvarez/STEALTH.
\end{abstract}

\pagestyle{plain}
\section{Introduction}

Within the current AI industry,
there are many available models-- not
all of which can be inspected. ``Model stores'' are cloud-based
services that charge a fee to use models hidden away behind a firewall (e.g. AWS market-place \cite{marketplace2019machine} and 
the Wolfram neural net repository\cite{wolfram}). Adams et al. \cite{xiu2020exploratory} discusses model stores (also known as  ``machine learning as a service''\cite{ribeiro2015mlaas}), and warns that  these models  are often  low quality   (e.g. if it comes from  a hastily constructed prototype from a GitHub repository, dropped into a container, and then sold as a cloud-based service). Theoretically,  software testing can   defend us against  low-quality models, but  in practice, these models do  not come with verification results nor detailed
specifications-- so standard testing methods are unsure what to test.

An alternative to  testing is explanation algorithms that offer a high-level summary of     a model.  Unfortunately,  the more
we learn about  explanations,
the better we   get at making misleading explanations. As we shall see: 
\begin{formal} 
 ~~~~The  more a model is queried, the better it can mislead.
\end{formal}
For example, Slack et al.'s lying algorithm\cite{slack}
(discussed below) knows how to detect ``explanation-oriented'' queries, and can then switch to  models which, by design,   disguise  biases,  or unfairness,   against  marginalized groups (e.g. by gender, race, age, or other identifiers). 

Slack et al.'s results are particularly troubling.
An alarming number of commercially deployed models have discriminatory properties\cite{noble18,chakraborty2021bias}. For example, the (in)famous COMPAS model (described in Table~\ref{tab:dataset})   decides the likelihood of a criminal defendant reoffending. The model suffers from alarmingly high false positive rates for Black defendants than White defendants.  Noble's
book \textit{  Algorithms of Oppression} offers a long list of other models with
discriminatory properties\cite{noble18}.
Clearly, before we can trust models from the cloud, we need ways to assess their bias without being misled by malicious algorithms. 

This article reasons as follows.  If too many queries are the problem, then perhaps the solution is:
\begin{formal} ~~~~~~~~~~~~~~~~~~~~~~~~~~Ask fewer queries.\end{formal} 
 To test this idea, we built a new algorithm called
 {\IT} which recursively bi-clusters $N$  examples,  down to leaves of size $\sqrt{N}$. A model is then queried with 1 example per leaf. These queries generate ``labels''; i.e. decisions about each $\sqrt {N}$ example, and are used to build a \textit{  surrogate} model which can be used for predictions and explanations. 
The key point here is since {\IT} makes few queries, 
malicious algorithms can not detect its operation. Hence, they never know when to lie.
 Also, since we use our surrogate for explanations, those explanations cannot be misled by a malicious model.

However, this  assumes the surrogate (learned from very few samples) can  mimic the important properties
of the original model(learned from more data). To test that, our research questions ask what is \textit{  gained} and \textit{  lost} by reasoning over such a small $\sqrt{N}$ sample of the data:

{\textbf RQ1:} \textit{  Does our method prevent lying?} 
When querying Slack et al.'s lying algorithm, we found while the original model can lie, \textit{  lying fails in our surrogates}.

{\textbf RQ2:} \textit{  Does the surrogate model perform as well as the original model?}
Much to our surprise, in the usual case, \textit{  {\IT}'s surrogates performed as well as or better than the original (\textit{MODEL1})}, measured in terms of \textit{  both}  predictive performance \textit{  and} fairness.  

{\textbf RQ3:} \textit{  How does {\IT} compare against
other bias mitigation algorithms?} Another reason to use STEALTH
is its fairness properties are competitive to those generated via state-of-the-art  bias mitigation methods such as Fair-Smote\cite{chakraborty2021bias}, MAAT\cite{chen2022maat}, and FairMASK\cite{peng2021xfair}.    
 \section{Background (How to Lie)}

\begin{figure}[!t]

\small
\begin{center}
\textbf{ Figure~\ref{fool}.A}:  LIME samples around the boundary to
find the delta between blue and red classes. From\cite{Ribeiro16}.
   
\includegraphics[width=.6\linewidth]{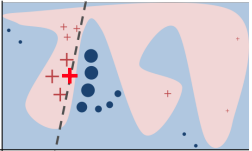} 

\textbf {Figure~\ref{fool}.B}: Feature importance as assessed by LIME. A \textbf{positive} weight means the feature encourages the classifier to predict the instance as a positive
and vice versa for the \textbf{negative} weight. Larger weights indicate greater feature importance.

\includegraphics[width=.8\linewidth]{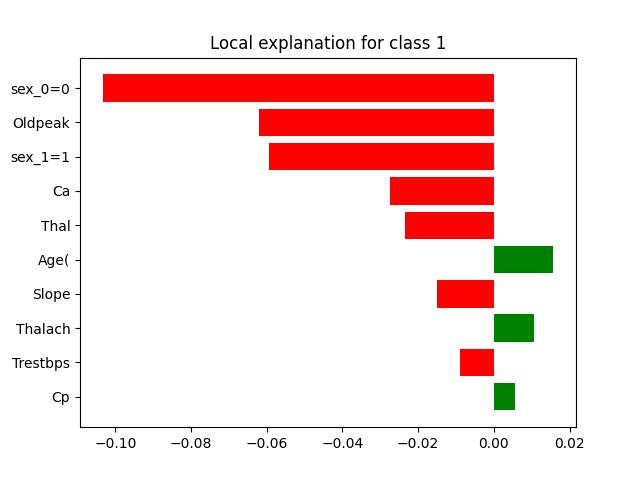}

\textbf{ Figure~\ref{fool}.C}: Blue/red = original/invented data \cite{slack}. 
 
\includegraphics[width=2.2in]{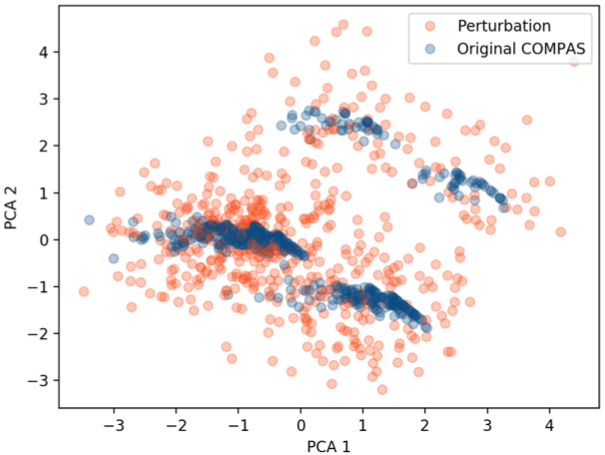}
 
\caption{About LIME.}
\label{fool}
\end{center}
\end{figure}

 Figure~\ref{fool}.A shows the Local Interpretable Model-Agnostic Explanations (LIME) algorithm\cite{Ribeiro16}  that  samples thousands of instances uniformly at random near an example point,  then builds a class of linear models from the generated sample data.  Using its models, LIME can explain what features influence the classification of a particular instance.

LIME's explanations look like   Figure~\ref{fool}.B. Note that, in the figure, the Age feature is most influential. By running LIME on all the test instances, it is possible to collect the \textit{  most influential}  features; i.e. the attributes that are found as being most influential for the test set.  

An interesting result seen in the most influential feature set is that often most features are \textit{  not} influential. For example, in our study, we have ten data sets with dozens of attributes, but only a handful are ranked most influential. 
This indicates the most influential features are a useful way to summarize the differences between explanations generated via different methods. Given the two most influential sets $A, B$  a Jaccard coefficient is computed between (0,1) meaning (zero, total) similarities in the explanations, respectively:
\begin{equation}\label{jacc}
\mathit{jaccard}(A, B) = \frac{|A \bigcap B|}{ |A\bigcup B|}
\end{equation}
 To build a lying algorithm, we note that, in 
Figure~\ref{fool}.C, many of    the  
{{\textbf red}} dots (showing LIME's generated samples) are  at different locations to 
 the   raw data
 ({{\textbf blue}} dots).  
Slack et al.\cite{slack} built a classifier to recognize when many of these {{\textbf red}} dots appear as input to the model. 
This meant  they could
   distinguish 
(a)~queries-about-explanation from
  (b)~normal-queries  seen  in  day-to-day  operation.
  Hence they could hide any discriminatory properties
  by   
  offering a \textit{  lying model}  to
  queries-about-explanation and   a second  model to normal-queries.

To build their liar, Slack et al. trained an out-of-distribution (OOD) model using similar mutations to LIME and used this OOD detector to produce different classifications using the \textit{protected attributes}.  These are special attributes that  divide data into two,  
\textit {privileged} \& \textit {unprivileged}, groups (e.g.
\textit {male} \& \textit {female}; \textit {young} \& \textit{old}). Thus, by training a model to recognize OOD data and in-distribution data, it is possible to hide discriminatory classifications against \textit {unprivileged groups}, and mislead explanations.

The problems raised by Slack et al.\cite{slack} are widely  explored-- see Table~\ref{related}. In that table,
{\IT}'s use of limited domain data as samples is novel.
Are such small samples of domain data relevant?
We claim, without controversy, that for a model store to profit, \textit{  consumers must want to use it}, and know their own domain data they wish to label. For example, a doctor could have  many records of   patients and want to query a model to obtain a label of ``high risk'' or ``low risk'' of an imminent heart attack. More formally,
 given a data set with  independent  attributes $X$ and dependent classes $Y$, our approach assumes access to many $x_i$ rather than other researchers who assume no access to $x_i$ or $y_i$ and instead use randomly generated $x_i$ values. The starting point for this research was the realization that often model users do have $x_i$ data, but not $y_i$ labels. Our goal is to learn $Y=\mathit{MODEL1}(X)$ under the following ``{\IT} assumptions'':

%   \[
% \begin{array}{rccc}
%     & X&& Y \\ 
% \mathit{eg}_1: &   (x_1,x_2,...)   &\rightarrow &  (y_1,y_2,...)  \\
% \mathit{eg}_2: & ...                                       &\rightarrow & ... \\
% \mathit{eg}_3: & etc\\
% \end{array}\]

\begin{itemize}
    \item We have domain data for   $x_i$, but not $y_i$
    \item We can obtain $y_i$ values by querying a model \textit{  MODEL2}
    \item \textit{  MODEL2} may be malicious
    \item To avoid being misled, we must limit the number of  $y_i$  queries
\end{itemize}

Given the {\IT} assumptions,  we first cluster our data on the $x_i$ values and then make one $y_i$ query per cluster.
Using the few $x_i$ and associated $y_i$ labels, we build a \textit{  surrogate} model.  
The limited $y_i$ queries would add very few {{\textbf red}} points 
to Figure~\ref{fool}, because the points are domain data samples and would be identical to the {{\textbf blue}} dots making our {\IT} approach inherently undetectable.

\begin{table*}
\caption{Related Work}\label{related}
\small
\begin{tabular}{|p{.99\linewidth}|}\hline
If the problem is that, in Figure~\ref{fool}.C, the {{\textbf red}} dots are different to the {{\textbf blue}} then one solution
is to make LIME's perturbations more realistic by adjusting sampling distributions~[1], or better neighborhood calculations [2,3,4,5].
Other approaches   improve LIME by making it more robust and less vulnerable to attacks~[4,5], or creating better detection defenses~[6,7] This is a very  active area of research see [8-12] where  prior work is rapidly assessed, improved, or refuted by subsequent work.

~~~~~We have three comments on that related work.
Firstly, some of that work makes restrictive assumptions about the data.
For example, Ji et al.~[1]   rely on 
parametric assumptions such as the data and model output conforming to a normal distribution.   While such assumptions might
be  true, they cannot be checked in the black-box case. {\IT}, on the other hand, is a non-parametric instance-based approach that makes no assumptions of normality.
 
~~~~~Secondly, some works assume complexity
when that may not be the best engineering decision for all domains.
As shown by the results of this paper,
methods that require 1000s of random samples (e.g. LIME) and which assume
very high dimensional data (e.g. generative adversarial networks [4,5]) are not appropriate in all domains.

~~~~~Thirdly, the above work does not take full advantage of non-generated data. Model store consumers
must have a supply of their own domain data they wish to label. We can use that to great effect as discussed in the main text. 

\vspace{2mm}
{\textbf Related work   citations:}\scriptsize
\begin{enumerate}
\item
  D. Ji, P. Smyth, and M. Steyvers, “Can i trust my fairness metric? assessing fairness with unlabeled data and bayesian inference,” in  NeurIPS'20.
\item
D. Vreˇs and M. R. ˇSikonja, “Better sampling in explanation methods can prevent dieselgate-like deception,”  arXiv:2101.11702, 2021.
\item
Y. Jia, J. Bailey, et al. “Improving the quality of explanations with local embedding perturbations,” in
KDD 2019, pp. 875–884.
\item
S. Saito, E. Chua, N. Capel, and R. Hu, “Improving lime robustness with smarter locality sampling,” arXiv preprint arXiv:2006.12302, 2020.
\item
A. Saini and R. Prasad, “Select wisely and explain: Active learning and probabilistic local post-hoc explainability,”  
AAAI/ACM Conference on AI, Ethics, and Society, 2022, pp. 599–608.
\item Z. Carmichael and W. J. Scheirer, “Unfooling perturbation-based post hoc explainers,” arXiv preprint arXiv:2205.14772, 2022.
\item J. Schneider, J. Handali, M. Vlachos, and C. Meske, “Deceptive ai explanations: Creation and detection,” arXiv preprint arXiv:2001.07641, 2020
\item Slack, D., Hilgard, A., Lakkaraju, H., \& Singh, S., "Counterfactual explanations can be manipulated," in NeurIPS '21
\item Wilking, R., Jakobs, M., \& Morik, K., "Fooling Perturbation-Based Explainability Methods," In ECML/PKDD '22.
\item Molnar, C., König, G., Herbinger, J., Freiesleben, T., Dandl, S., Scholbeck, C. A., ... \& Bischl, B., "General pitfalls of model-agnostic interpretation methods for machine learning models,". xxAI - Beyond Explainable AI. 2020.
\item Maratea, A., \& Ferone, A., "Pitfalls of local explainability in complex black-box models," In WILF. 2021.
\item Covert, I., et al. "Explaining by Removing: A Unified Framework for Model Explanation," J. Mach. Learn. Res. 22 (2021): 209-1.

\end{enumerate} \\\hline
\end{tabular}
\end{table*}
 
\begin{table*}
\caption{Inside {\IT}: The core of {\IT} is steps 2-5. Step 1,6, and 7 are added for evaluation. }
\label{rig}
\footnotesize%
\begin{tabular}{|p{\linewidth}|}\hline
%  \rowcolor{blue!10}
{\textbf 0. Data Preparation}
The available  data  is divided 40:40:20 into a Train1:Train2:Test split. 
The test set is used as a set of hold-out examples in steps 1, 6, and 7.
The Train1 set is used to build \textit{  MODEL1} that we are trying to understand (and which might be managed by a malicious model manager).
{\IT} does not use the class attribute labels in Train2 since its mission is to extract these from the limited  number of queries to \textit{  MODEL1} learned from Train1.
Hence, for these experiments,
all the class labels in Train2 are initially set to null. 
\\
{\textbf 1. Baseline generation:}
Build \textit{  MODEL1} (also known as the \textit{  original model})  using 40\% training data. In terms of our domain modeling, MODEL1 is the model hiding behind a firewall. Using MODEL1 and the 20\% hold-out test data,  we can collect \textit{  baseline values} for all the metrics of Table~\ref{metrics}.
% \\ \rowcolor{blue!10}
\\
{\textbf 2. Clustering:}
A  \textit{  clustering algorithm}    recursively bi-cluster Train2 data   to leaves of size $\sqrt{N}$. We use Figure~\ref{inside}'s recursive bi-clustering method. Nair et al. report that this method is fast and useful for finding a good spread of examples across   data\cite{chen2018sampling}.
\\
{\textbf 3. Sampling:} After that, we need a \textit{  leaf sampling algorithm} to select $m$ examples per leaf. Here, we selected $m=1$ items at random.
While {\IT} works at $m=1$, other algorithms such as MAAT\cite{chen2022maat}
seem to fail for such small data sets, and for some test cases, we cannot validate them with respect to prior work.\\
% \rowcolor{blue!10}
{\textbf 4. Labeling:} A  \textit{  labeling algorithm} must  then  assign labels to $\sqrt{N}$ selected examples. For that purpose, we query \textit{  MODEL1}.
\\
{\textbf 5. Surrogate Creation:} Some \textit{    model creation algorithm} must  build a surrogate model.  Here again,  we used a random forest to build \textit{  MODEL2} (also known as the \textit{  surrogate model}) using the  data
associated with the $\sqrt{N}$ labels. 
\\
% \rowcolor{blue!10}
{\textbf 6. Collect Performance Scores:} 
Here, the surrogate MODEL2 is applied to the 20\% hold-out test
data to generate the Table~\ref{metrics} performance scores.  These values are compared to
the {\textit{  baseline values}  seen in {\textbf step 1.}}
\\
 {\textbf 7. Collect Explanation Scores:} 
Here, we gather the most influential attributes of (b1)~MODEL1 and (b2)~MODEL2 using LIME, and 
compare them using Equation~\ref{jacc}.\\\hline
\end{tabular}
~\\

\end{table*}

\begin{figure*}
\caption{\label{inside}{\IT}  partitions   $x$ values using the   random projections. Nair et al. report that this method is fast and is useful for generating a good spread of examples across a data set [10].} 
 \small~\\
\begin{minipage}{2in}%
\includegraphics[height=3.6cm]{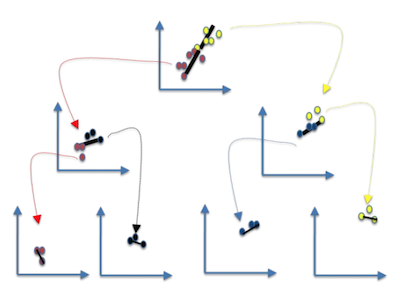}%
\end{minipage}%
\begin{minipage}{1.7in}%
\includegraphics[height=3.6cm]{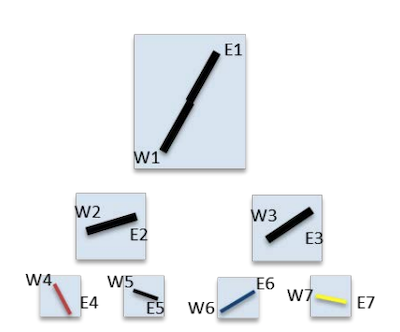}\end{minipage}\hspace{3mm}%
 \begin{minipage}{3.5in}\small
 \textbf{NOTES:}  This recursive bi-clustering
 method finds two distant points  
  $E$  (east) and 
  $W$ (west).
  Using  
  \mbox{$c=\mathit{dist}(E,W)$} then
    all other examples have distances $a,b$  to $E,W$, respectively and distance
    \mbox{$x=(a^2+c^2-b^2) / (2c)$}
    on a line from  $E$ to $W$.
  By
splitting   data on   median $x$,
the examples can be then bi-clustered
(and so on, recursively, see   Figure~\ref{inside}b).
\end{minipage}  

\begin{minipage}{2in}{\textbf{\hspace{.6in}Figure~\ref{inside}a.}}\end{minipage}\hspace{3mm}\begin{minipage}{2in}\hspace{.5in}{\textbf{ Figure~\ref{inside}b.}}\end{minipage}\begin{minipage}{2in}~\end{minipage}\hspace{3mm}%

\end{figure*} 
 
\begin{table*}[!t]
\caption{Evaluation metrics. In this table, TP, TN, FP, FN are the true/false positive/negative rates seen in binary classification. The definition for True Positive Rates (TPR) is TP$/$(TP $+$ FN) and  False Positive Rates (FPR) is FP$/$(FP $+$ TN).}
\scriptsize
\tabcolsep=0.11cm
\begin{center}
\begin{tabular}{|l|l|}
\hline
\multicolumn{1}{|c|}{ \textbf{Performance Metric}} & \multicolumn{1}{c|}{ \textbf{Fairness Metric}}   \\ \hline
% \rowcolor{blue!10}
Accuracy = (TP$+$TN)$/$(TP$+$TN$+$FP$+$FN) &
\begin{tabular}[c]{@{}l@{}}\textbf{Average Odds Difference (AOD)}: Average of difference in False Positive Rates(FPR) and \\ True Positive Rates(TPR) for unprivileged and privileged groups. \\ $AOD$ $=$ (($FPR_{U} $-$ FPR_{P}$) + ($TPR_{U} $-$ TPR_{P}$))$/$ 2 \end{tabular}  \\ \hline

False alarm = FP/(FP+TN)  & \begin{tabular}[c]{@{}l@{}}\textbf{Equal Opportunity Difference (EOD)}:  Difference of True Positive Rates(TPR) for \\ unprivileged and privileged groups.   $EOD = TPR_{U} - TPR_{P}$\end{tabular}  \\ \hline
% \rowcolor{blue!10}
Recall = TP/(TP+FN)  & 
\begin{tabular}[c]{@{}l@{}}\textbf{Statisticl Parity Difference (SPD)}: Difference between probability  of unprivileged \\ group (protected attribute PA = 0) gets favorable prediction ($\hat{Y} = 1$) \& probability \\ of privileged group (protected attribute PA = 1) gets favorable prediction ($\hat{Y} = 1$). \\ $SPD = P[\hat{Y}=1|PA=0] - P[\hat{Y}=1|PA=1]$\end{tabular}     
\\ \hline

Precision =  TP$/$(TP$+$FP) & 
\begin{tabular}[c]{@{}l@{}}\textbf{Disparate Impact (DI)}: Similar to SPD but instead of the difference of probabilities, \\ the ratio is measured. $DI = P[\hat{Y}=1|PA=0]/P[\hat{Y}=1|PA=1]$\end{tabular}   
\\ \hline
% \rowcolor{blue!10}
 F1 Score = 2 $\times$ (Precision $\times$ Recall)$/$(Precision $+$ Recall) &  \\ \hline

\end{tabular}
\end{center}

\label{metrics}
 \end{table*}

\begin{table*}[!t]
\caption{Description of processed data sets including their domain usage, protected attributes, numerical features post-processing, and non-empty rows. \IT{} number of rows and median run-times in seconds over 20 repeats.} \label{tab:dataset}
\begin{framed}
\scriptsize
\tabcolsep=0.11cm
 \begin{center}
\begin{tabular}{rccrrrrr}
  &    &    &   &  &  &   Our \\ 
  &    & Protected  & Num & N & $\sqrt{N}$ &   runtimes \\ 
Data set &   \multicolumn{1}{c}{Domain} &    Attribute & Features & rows & rows & (secs) \\ \hline
Adult Census~[1]  & U.S. census information from 1994 to predict personal income & Sex, Race & 7  & 45,522 & 512 & 58.89 \\
Bank Marketing~[2]  & Marketing data of a Portuguese bank to predict term deposit  & Age & 7 & 30,488 & 512 & 54.54     \\
Default Credit~[3]  & Customer information in Taiwan to predict default payment & Sex & 24   & 30,000 & 512 & 94.98 \\
Compas~[4]         & Criminal history of defendants to predict re-offending  & Sex, Race & 4 & 6,172 & 256 & 12.43 \\
MEPS15~[5]        & Surveys of household members and their medical providers & Race & 41  & 4,870   & 128   & 2.00 \\
Student~[6]        &  Student performance to predict good grades & Sex & 24  & 1,044 &  64  & 1.19 \\
German Credit~[7]  & Personal information to predict good or bad credit & Sex & 6    & 1,000  &  64    &  1.19\\
Diabetes~[8]  & Diagnostic measurements to predict diabetes & Age & 8    & 768 & 64   &  1.14 \\
Heart Health~[9]   & Patient information from Cleveland DB to predict heart disease & Age & 13    & 297 & 58    & 1.11 \\

Communities ~[10]  &  Law enforcement information to predict violent crimes & racePctWhite & 122    & 123 &  32  & 1.49\\
\end{tabular} 
\end{center}
 
\textbf{ Data set citations:}
\begin{enumerate}

\item "Adult data set,” 1994.  \textit{  http://mlr.cs.umass.edu/ml/datasets/Adult}
\item “Bank marketing,” 2017.    \textit{   https://www.kaggle.com/c/bank-marketing-uci}
\item  “Default of credit card clients data set,” 2016.    \textit{   https://archive.ics.uci.edu/ml/datasets/default+of+credit+card+clients}
\item  “Compas Analysis,” 2015.    \textit{   https://github.com/propublica/compas-analysis}
\item  “Medical expenditure panel survey,” 2015.    \textit{   https://meps.ahrq.gov/mepsweb/}
\item  “Student performance data set,” 2014. 
  \textit{   https://archive.ics.uci.edu/ml/datasets/Student$+$Performance}
\item  “German credit data set,” 2000.    \textit{   https://archive.ics.uci.edu/ml/datasets/Statlog+German+Credit+Data}
\item  U. M. Learning, “Pima indians diabetes database,” 2016.  \textit{   https://kaggle.com/uciml/pima-indians-diabetes-database}
\item  “Heart disease data set,” 2001.  \textit{   https://archive.ics.uci.edu/ml/datasets/Heart+Disease}
\item M. Redmond, “Communities and crime unnormalized data set,” 2011.
 \textit{   http://www.ics.uci.edu/mlearn/ML-Repository}
\end{enumerate}
\end{framed}

\end{table*}

\section{Methods}
Table~\ref{rig} describes the seven steps needed to run and evaluate {\IT}.
To answer our research questions, we applied those steps, with and without Slack et al.'s lying algorithm \cite{slack}. 
The surrogate models built were assessed via their (a)~LIME explanation properties (using Equation~\ref{jacc}); (b)~their predictive performance; and (b)~their fairness using metrics that are widely applied in the literature (see Table~\ref{metrics}). 
Since we are exploring fairness, we also compare {\IT} to state-of-the-art bias reduction algorithms Fair-Smote\cite{chakraborty2021bias}, MAAT\cite{chen2022maat}, and FairMASK\cite{peng2021xfair}.

\textbf{ALGORITHMS:} For this work, we tried various learners such as random forests, logistic regression, and a SVM with a radial basis function.
It was observed that random forests generated better recalls, and thus, we focus on random forests for this work. 
 
 Table~\ref{rig} uses several bias reduction methods. 
Chen et al.'s MAAT system\cite{chen2022maat}  makes conclusions by balancing   conclusions between   two separate models  (a performance model and a fairness. model).
Chakraborty et al.'s Fair-SMOTE system\cite{chakraborty2021bias} adjusts the training data by (a)~removing training instances that change conclusions when protected attributes change; and (b)~evening out the distributions of the protected attribute values.  Chakraborty argues that these adjustments  make it harder for any particular protected value to unduly affect the conclusion. 
Peng et al.'s FairMASK system\cite{peng2021xfair} replaces protected attributes with new values learned   from other independent attributes. Peng argues that this removed misleading latent correlations between attributes since the learned values tend to be average values (not outliers).

 An important note to make about MAAT, Fair-SMOTE,  and FairMASK is that all these methods require the analyst to pre-specify the protected attributes. {\IT}, on the other hand, operates without such knowledge. Despite this, as we will see, {\IT} achieves competitive levels
 of bias reduction to MAAT, Fair-SMOTE,  and FairMASK.

\textbf{DATA:}
The  case studies in
Table~\ref{tab:dataset}
 were selected since they appeared in prior  explanation and fairness papers including Slack et al. \cite{slack}.
In column three of Table~\ref{tab:dataset}, each
protected attribute is listed.

{\textbf STATISTICS:} Eight of our data sets have one protected attribute while  ADULT and COMPAS have two. Hence, we run 12 times, one for each data set and protected attribute. Each run is repeated 20 times, each time dividing the available data into 40\% Train1, 40\% Train2, and 20\% Test.
Since each run returns nine scores (one for each of the metrics of Table~\ref{metrics}), then generates 12*20*9=2160 scores. 

For each data set and metric,
we compared   methods 
 (using {\IT} or MAAT or FairSMOTE or FairMASK) against a {\textbf baseline}
(running all  data through random forests). 
Each method earns one more win, tie, or loss 
if it is statistically better, the same, or worse
than the {\textbf baseline} (respectively). 
For that statistical test, we use a nonparametric test recommended by prior work on bias reduction; a Scott-Knot procedure using Cliffs Delta
and a bootstrap test. For
notes justifying and explaining those procedures, see\cite{hess2004robust,mittas2012ranking,chakraborty2021bias,chen2022maat}.
Note that  for recall, precision, accuracy, and F1 \textit{  larger} numbers are \textit{  better} while for all other metrics such as false alarms, \textit{  smaller} numbers are \textit{  better}.

 \section{Results} \label{Results}
 \textit{  {\textbf RQ1:}  Does our method prevent lying?}

  Table~\ref{jac} compares the explanations from Slack et al.'s lying algorithm and  {\IT}'s Slack surrogate models. In Table~\ref{jac}, we see that the Slack et al. entries are nearly all zero; i.e. explanations from {\IT} do not  overlap with the lying algorithm's explanations. This
  means  that Slack's algorithm cannot detect the operation of {\IT}, and so it never knows when to lie.
  
  Generating different explanations, from the liar, could be achieved by selecting lucky random values. Happily, we can show that the explanations from {\IT} correspond to
  important influences within our models.   Table~\ref{jac}
  also shows us that 
explanations from {\IT} have a large overlap (median=59.5\%) with the explanations from the original model, \textit{MODEL1}.  
 
In summary, to answer RQ1, we say {\IT} can defeat Slack et al.'s lying algorithm. This is not to say that {\IT} defeats all possible liars, and exploring our method in the context of other liars is a clear direction for future work.

For our methods, RQ1 is a positive result while
many of the explanation overlaps reported in  Table~\ref{jac} are  less than 100\%. That is, between the original model and those made by {\IT}, there are differences measured in terms of which  attributes are most influential. Happily, in conjunction with the results of RQ2, it suggests {\IT}'s models are preferable to the original, and arguably we should have more confidence in the most influential attributes, as reported by {\IT}. 

\begin{table}[!t]
\caption{RQ1 results: Jaccard coefficient (Equation~\ref{jacc}) results comparing LIME's first top-ranked features of \textit{MODEL1} (with the Baseline and the Slack et al. liar algorithm) to \IT{}'s first top-ranked features.}
\label{jac}
\scriptsize
\centering
\begin{tabular}{|l|l|l|}
\hline
Data set & Slack\_Jacc & Base\_Jacc  \\ \hline
student & 0.08 & 0.25 \\
meps & 1 & 0.42 \\
adult\_r & 0 & 0.50 \\
adult\_s & 0 & 0.50 \\ 
compas\_s & 0 & 0.57\\
 &   & 0.595 $\Longleftarrow$ median\\
diabetes & 0.20 & 0.62 \\
compas\_r & 0.0 & 0.67 \\
default & 0.07 & 0.68 \\
bank & 0.17 &  0.69\\
german & 0 & 0.79 \\
\hline
\end{tabular}%
\end{table}

\begin{table*}[!t]
\caption{RQ2 results: A method earns one win or tie or loss if it is statistically 
{\textbf better}, same, or worse(respectively) than the Figure~\ref{basenum} baseline (as assessed via Scott-Knot statistical tests that combine   Cliffs Delta
and a bootstrap). Note that  for recall, precision, accuracy, and F1 \textit{  larger} numbers are \textit{  better} while for all other metrics such as false alarms, \textit{  smaller} numbers are \textit{  better}.
}
\label{compare}
\scriptsize
\begin{center}
\begin{tabular}{|l|lllc|lllc|}
\hline
 & \multicolumn{4}{|c|}{Performance:} 
 &  \multicolumn{4}{|c|}{Fairness:} \\
 & \multicolumn{4}{|c|}{(accuracy, recall, precision, F1, false alarm)} & 
  \multicolumn{4}{|c|}{(AOD, EOD, SPD, DI)}\\ 
          &     &       &    &           &      &     &      &\\
Method & Wins & Loses & Ties &  Wins + Ties & Wins & Loses & Ties &  Wins + Ties \\ \hline
STEALTH & 28 & 7 & 25 & 53 / 60 & 21 & 6 & 21 & 42 / 48\\
MAAT & 34 & 17 & 9 & 43 / 60  & 31 & 7 & 10 & 41 / 48 \\
Fair-SMOTE & 27 & 17 & 16 & 43 / 60 & 28 & 6 & 14 & 42 / 48 \\
FairMask & 21 & 23 & 16 & 37 / 60  & 12 & 25 & 11 & 23 / 48\\ \hline
\end{tabular} 
\end{center}
\end{table*}

\textit{  {\textbf RQ2:}   Does the surrogate model perform as well as the  original model?}

 It turns out that our method for avoiding liars does not lead to sub-optimal models.
Table~\ref{compare} shows how often our method finds improvements over the {\textbf baseline}  models. Note that there are many wins recorded in   Table~\ref{compare} such as achieving numerous non-small, statistically significant improvements in performance and fairness compared to the baseline.  
 
\begin{figure}[!t]
\begin{center}
\footnotesize
\includegraphics[width=3.5in]{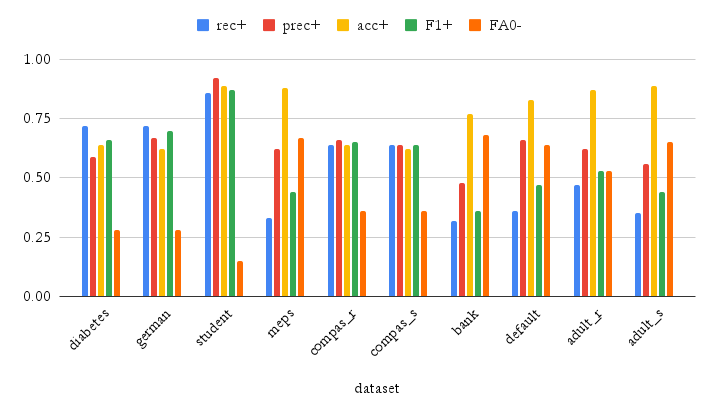}
\vspace{3mm}
 Figure~\ref{basenum}.A: 
 RQ2 results: \textit{  performance} reports median results for accuracy, recall, false alarm precision, F1.
\vspace{3mm}
\includegraphics[width=3.5in]{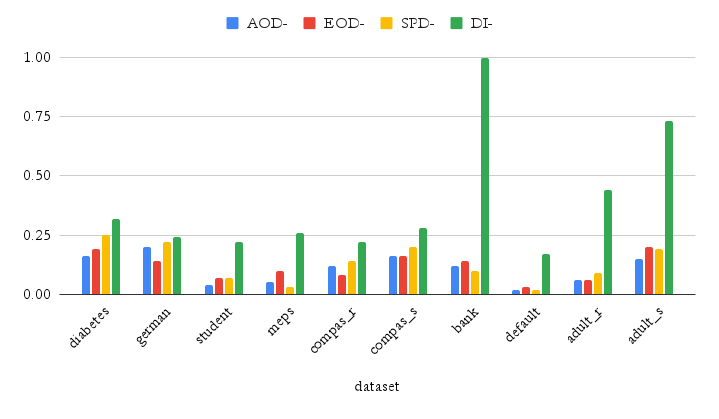}
\vspace{3mm}
 Figure~\ref{basenum}.B: RQ2 results: \textit{  fairness} reports median results for EOD, AOD, SPD, DI
\vspace{3mm}
\end{center}
\caption{RQ2 {\textbf Baseline}, a non-malicious RF trained with all data,   results
on Table~\ref{metrics} metrics.}\label{basenum}
\end{figure}

 There are some losses reported in Table~\ref{compare} and to explain, we look at the raw values of the {\textbf baseline} models (shown in Figure~\ref{basenum}). It is clear that in the original model, \textit{MODEL1}, there are  some weakly performing data sets.  For example, in Figure~\ref{basenum}, while ADULT has high accuracy, the baseline also suffers from very high false alarms (over 50\%). This is relevant since, when we drill down into the losses of Table~\ref{compare}, we note that many of {\IT}'s losses arise in  ADULT.
That is to say, these losses seem to be less about drawbacks with {\IT} and rather drawbacks with random forests (for these data sets).
 
\textit{  {\textbf RQ3:}   Does the surrogate model enhance fairness? How does it compare to SOTA fairness mitigation methods?}

Our summary of Table~\ref{compare} is that {\IT} is competitive with the current state-of-art, and even defeats some of those algorithms (e.g. FairMask).
A case might be made that Fair-SMOTE and MAAT perform a little  better since they have a few more wins
(that said, the wins+ties are the  same as {\IT}).

Hence, we see no reason to dismiss {\IT} based on the fairness of the models it creates, and given its superior \textit{  performance} scores, makes us recommend this algorithm.

Another reason to recommend {\IT} is that
  other bias reduction methods have to be directed to protect specific attributes. {\IT},
  on the other hand, offers bias improvements without having to be directed. This is to say {\IT}  could protect important attributes even if they were inadvertently overlooked.

\section{Discussion: Why Does it Work So Well?} 
By using a few examples, {\IT} runs ``under the radar'' and is undetected by Slack et al.'s lying algorithm. 
{\IT} makes very few ($\sqrt{N}$) in-distribution queries,  which are undetectable by OOD attacks. Since {\IT}'s queries are undetectable, the liar cannot lie and returns honest results.
Why do  models built from just  $\sqrt{N}$ examples have similar or better performance, and bias reduction properties?
Perhaps this is for the same reason that semi-supervised reasoning\cite{mit06} works so well.
One of the relevant semi-supervised learner assumptions is the ''manifold assumption", i.e. higher dimensional data can be approximated by a smaller set of most important attributes, without much loss of signal \cite{mit06}.  As evidence of our data having a lower dimensional manifold, recall from the above that in all the data sets used here, there were  very few most influential attributes.  
When such smaller manifolds exist,
data mining needs to only explore a small sample of the data set (provided the samples are spread across the data set using methods like, for example, Figure~\ref{inside}). 

 As to {\IT}'s success at reducing bias, 
we  conjecture that {\IT}'s success results from a  zealous  use of design principles from Peng et al.'s FairMASK \cite{peng2021xfair}. Peng et al. argue that FairMASK reduces bias by replacing minority outlier opinions with values that are better supported by more  data. FairMASK does this by synthesizing replacement values for protected
attributes from the other independent attributes.
{\IT}, on the other hand, does this for \textit{  all independent
attributes} since, in effect, its $\sqrt{N}$ examples are reports of average case effects
within small data clusters. By averaging the data in this way,  we \textit{smooth} out the features including any outliers that may skew results unfairly.

(\textbf{Technical aside}: other fairness mitigation algorithms e,g, Fair-SMOTE, mitigate for unfairness by ``blanding-out''   attributes-- i.e. by giving their values equal frequency in the training data. Our pre-experimental concern was that {\IT} operations on all attributes would bland-out
everything, thus leading to a naive and poorly performing classifier.
Yet, we realized it was the opposite of blanding-out.
 {\IT}  builds a training set by clustering down to $\sqrt{N}$ and uses a random point near the centroids of the leaf clusters to build a surrogate. In so doing, {\IT}
 removes  outlier signals
and emphasizes the central tendencies within each of these $\sqrt{N}$ samples. Hence, we achieve the \textit{  fairness} results of Table~\ref{compare} by deleting spurious signals without having to explicitly annotate which attributes are protected.)

Based on the above,    {\IT}  can be called a fairness enhancing algorithm    
which we  recommended over MAAT, Fair-SMOTE, and FairMASK. 
But also,  we argue that there
is another result here. 
An important feature of these results is they
suggest a new view  on the  nature of explanation, discussion, and discrimination.   The {\IT} results
highlight  a connection between ``better, more honest explanations'' and ``reducing biased results'' into something  that we might call ``trusted communication''.
  Are there other such ``trusted communication'' methods? Would they perform better than {\IT}? We do not know-- but that would be a useful direction for future research.

\section*{Acknowledgements}
This work was partially funded with support from IBM and the National Science Foundation Grant \#1908762.
Any opinions, findings, conclusions, or recommendations
expressed in this material are those of the authors and do
not necessarily reflect the views of IBM or the NSF.

\small
% \balance
\bibliographystyle{IEEEtran}
\bibliography{references}

\end{document}